\documentclass[amsmath, amssymb, superscriptaddress, 
twocolumn, pra]{revtex4-1}

\usepackage{graphicx}
\usepackage{color}
\usepackage{hyperref}
\usepackage{CJK}

\begin{document}
	
\title{Gain/loss effects on spin-orbit coupled ultracold atoms in two-dimensional optical lattices}

\begin{CJK*}{UTF8}{gbsn}
\author{Zhi-Cong Xu (许志聪)}
\thanks{They contribute equally to this work}
\author{Ziyu Zhou (周子榆)}
\thanks{They contribute equally to this work}
\author{Enhong Cheng (成恩宏)}
\author{Li-Jun Lang (郎利君)}
\email{ljlang@scnu.edu.cn}
\affiliation{Guangdong Provincial Key Laboratory of Quantum Engineering and Quantum Materials, School of Physics and Telecommunication Engineering, South China Normal University, Guangzhou 510006, China}
\author{Shi-Liang Zhu (朱诗亮)}
\affiliation{Guangdong Provincial Key Laboratory of Quantum Engineering and Quantum Materials, School of Physics and Telecommunication Engineering, South China Normal University, Guangzhou 510006, China}
\affiliation{Guangdong-Hong Kong Joint Laboratory of Quantum Matter, Frontier Research Institute for Physics, South China Normal University, Guangzhou 510006, China}

\date{\today}

\begin{abstract}
Due to the fundamental position of spin-orbit coupled ultracold atoms in the simulation of topological insulators, the gain/loss effects on these systems should be evaluated when considering the measurement or the coupling to the environment. 
Here, incorporating the mature gain/loss techniques into the experimentally realized spin-orbit coupled ultracold atoms in two-dimensional optical lattices, we investigate the corresponding non-Hermitian tight-binding model and evaluate the gain/loss effects on various properties of the system, revealing the interplay of the non-Hermiticity and the spin-orbit coupling.
Under periodic boundary conditions, we analytically obtain the topological phase diagram, which undergoes a non-Hermitian gapless interval instead of a point that the Hermitian counterpart encounters for a topological phase transition. We also unveil that the band inversion is just a necessary but not sufficient condition for a topological phase in two-level spin-orbit coupled non-Hermitian systems.
Because the nodal loops of the upper or lower two dressed bands of the Hermitian counterpart can be split into exceptional loops in this non-Hermitian model, a gauge-independent Wilson-loop method is developed for numerically calculating the Chern number of multiple degenerate complex bands. 
Under open boundary conditions, we find that the conventional bulk-boundary correspondence does not break down with only on-site gain/loss due to the lack of non-Hermitian skin effect, but the dissipation of chiral edge states depends on the boundary selection, which may be used in the control of edge-state dynamics.
Given the technical accessibility of state-dependent atom loss, this model could be realized in current cold-atom experiments.
\end{abstract}

\maketitle
\end{CJK*}	

\section{Introduction}
Spin-orbit coupling is a key element to realize topological insulators in condensed matters \cite{BernevigHughes2013}, and to realize it is a basic task for each experimental platform that aims to simulate topological physics. 
Cold atoms as a quantum simulator are such a platform that has promising potential in solving problems of many-body systems, quantum computations, etc \cite{LewensteinSen2007,BlochZwerger2008}.
In 2011, Spielman's group first realized one-dimensional spin-orbit coupling with ultracold atoms \cite{LinSpielman2011}, and later the success has been extended to two-dimensional (2D) fermions \cite{HuangZhang2016,MengZhang2016} and bosons \cite{WuPan2016,SunPan2018}; these achievements pave the way for simulating topological matters via cold atoms \cite{ZhaiZhai2015,ZhangZhu2018}.

In realistic experiments, the loss cannot be completely avoided due to the coupling of systems to the environment or measurement \cite{BreuerPetruccione2002}; for cold atoms, few-body losses play inevitable roles in the preparation of degenerate quantum gases \cite{LewensteinSen2007} and in the simulation of quantum many-body physics \cite{BlochZwerger2008}. On the other hand, the non-Hermitian physics attracts increasing attention of almost all branches of physics in recent years \cite{AshidaUeda2020}, and abundant exotic phenomena have been widely exploited both in theory and experiment, such as the spontaneous breaking of parity-time ($\mathcal{PT}$) symmetry \cite{BenderBoettcher1998,GuoChristodoulides2009,PengYang2014,PoliSchomerus2015,LiLuo2019,TakasuTakahashi2020,DingZhang2021,RenJo2021}, the breakdown of conventional bulk-boundary correspondence \cite{LeeLee2016,LeykamNori2017,ShenFu2018,YaoWang2018,GongUeda2018,XiongXiong2018,KunstBergholtz2018,MartinezAlvarezFoaTorres2018,YinChen2018,JinSong2019,BorgniaSlager2020,ZhangFang2020}, the exceptional topology \cite{BergholtzKunst2021}, and the interplay with Anderson localization \cite{JiangChen2019,Longhi2019,Xu2020,ZhangZhu2020,XuChen2021,LiuChen2021,LinXue2021,TangZhang2021}. 
As for cold atoms, the experimental techniques are mature to engineer state-dependent atom losses \cite{LiLuo2019,LappGadway2019,GouYan2020,TakasuTakahashi2020,FerriEsslinger2021,DingZhang2021,RenJo2021} and the effective nonreciprocal hoppings \cite{GouYan2020} of non-Hermitian systems, which are fundamental operations for the construction of a non-Hermitian model. 

Since the non-Hermiticity can be experimentally engineered in cold atoms, we're wondering about gain/loss effects on spin-orbit coupled ultracold atoms in optical lattices.
To this aim, by incorporating the gain/loss techniques \cite{LiLuo2019,LappGadway2019,GouYan2020,TakasuTakahashi2020,FerriEsslinger2021,DingZhang2021,RenJo2021} into the spin-orbit coupled ultracold atoms in 2D optical lattices experimentally realized in Ref. \cite{WuPan2016,SunPan2018}, we investigate the corresponding four-band tight-binding model with both the spin-dependent and the sublattice-staggered gains/losses, and analytically illustrate a gain/loss-induced topological phase transition by the method of block diagonalization. 
{Different from the Hermitian counterpart of which the transition occurs at a gapless point determined by the band inversion, the transition here undergoes a non-Hermitian gapless interval, unveiling that the band inversion in the real part is just a {\it necessary but not sufficient} condition for a topological phase in two-level spin-orbit coupled non-Hermitian systems.}
For a fully complex-gapped phase, the Chern number can be determined by the block-diagonalized Hamiltonian.

Moreover, the nodal loops between upper or lower two dressed bands for Hermitian cases protected by nonsymmorphic symmetries \cite{LangZhou2017} are split into exceptional loops for non-Hermitian cases in the presence of a purely imaginary staggered potential. The existence of these exceptional loops motivates us to develop a Wilson-loop method for numerically calculating the Chern number of multiple degenerate complex bands, and we find that only with \textit{dual} left/right eigenvectors is the Chern number gauge-independent. This method can be regarded as a non-Hermitian generalization of the non-Abelian scheme in Hermitian systems \cite{FukuiSuzuki2005}.

At last, we demonstrate the preservation of conventional bulk-boundary correspondence {due to the lack of non-Hermitian skin effect}, but the dissipation of chiral edge states under open boundary conditions (OBCs) depends on the boundary selection, which may be used in the control of edge-state dynamics. 

This work deepens the understanding of gain/loss effects on topological insulators and of the interplay between non-Hermiticity and spin-orbit coupling, and may stimulate corresponding simulations with cold atoms as well as other experimental platforms, such as photonics \cite{ZeunerSzameit2015,PoliSchomerus2015,ZhuChen2018,XiaoXue2020,WeidemannSzameit2020,WangFan2020}, nitrogen-vacancy centers \cite{WuDu2019,ZhangDuan2021}, electrical circuits \cite{HelbigThomale2020,HofmannNeupert2020}, and mechanical systems \cite{BrandenbourgerCoulais2019,GhatakCoulais2020}.

\section{The non-Hermitian tight-binding model}
\begin{figure}[tb]
  \includegraphics[width=0.9\linewidth]{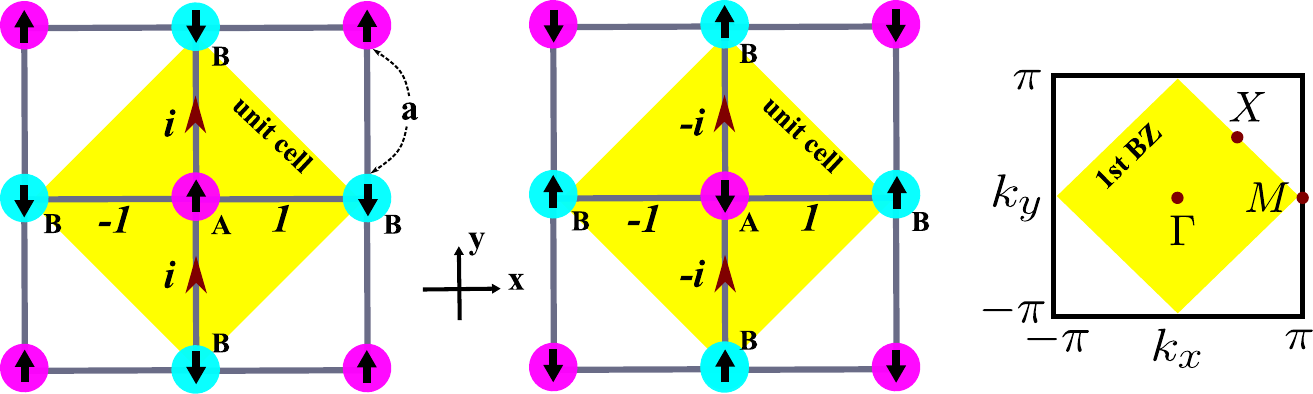}
  \caption{First two panels: schematics of the spin-orbit coupling term ($t'=1$) in Eq. \eqref{eq:Ham}. The square lattice is divided into two sublattices (A and B) and becomes a checkerboard pattern with a larger unit cell (yellow shaded squares). 
  Last panel: 1st BZ of the square lattice compared with the smaller one (yellow shaded square)  of the checkerboard lattice, where high symmetric points $\Gamma(0,0),X(\pi/2,\pi/2)$ and $M(0,\pi)$ are denoted ($a=1$).}
\label{fig1}
\end{figure}

The tight-binding Hamiltonian for spin-orbit coupled ultracold atoms in a square lattice \cite{WuPan2016,SunPan2018} with on-site gain/loss can be generally written as
\begin{eqnarray}
\hat{H}=\sum_{\sigma=\uparrow,\downarrow}
\Big\{&&
\sum_{\langle nm \rangle}[t\,\hat{\psi}_{n\sigma }^{\dag}\hat{\psi}_{m\sigma }
+(-1)^{n_x+n_y}t^{\prime }e^{i\eta_\sigma\phi_{m}}\hat{\psi}_{n\sigma }^\dag\hat{\psi}_{m\bar{\sigma} }] \notag \\
&&+\sum_{n}\left[\eta_\sigma h+(-1)^{n_x+n_y}\Delta\right]\hat{\psi}_{n\sigma }^\dag \hat{\psi}_{n\sigma}\Big\},
\label{eq:Ham}
\end{eqnarray}
where $n=(n_x,n_y)$ is a collective index of a site position $\mathbf{R}_n=a(n_x\hat{\mathbf{x}}+n_y \hat{\mathbf{y}})~(n_{x,y}\in \mathbb{Z})$ with the lattice constant $a$,
$\sigma$ and $\bar{\sigma}$ stand for a hyperfine spin and its spin-flip,
and $\hat{\psi}_{n\sigma }^{(\dagger) }$ is the annihilation (creation) operator for spin-$\sigma$ atom at site $n$.
As shown in Fig. \ref{fig1}, the second term represents a spin-orbit coupling with the inter-spin hopping strength $t'$ and the phase $\eta_\sigma\phi_m$ between nearest-neighbor sites denoted by $\langle nm\rangle$, where $\eta_{\sigma=\uparrow,\downarrow}=\pm$, and $\phi_{m}=0,-\pi/2,\pi, \pi/2$ for $m=(n_x+1,n_y),(n_x,n_y+1),(n_x-1,n_y),(n_x,n_y-1)$, respectively.
$h\,(\Delta)$ is a complex number and can be regarded as a complex Zeeman field (staggered potential), resulting from the state-dependent atom loss \cite{LiLuo2019,RenJo2021}.
In the following, we set the intra-spin hopping strength $t=1$ as the energy unit and $a=1$ as the length unit.

\section{Topological phase diagram}
Under periodic boundary conditions (PBCs), we take the discrete Fourier transform for a {\it square} lattice with $N$ sites,
$\hat{\psi}^\dag_{\mathbf{k}\sigma}=N^{-1/2}\sum_{n}e^{i\mathbf{k}\cdot \mathbf{R}_n}\hat{\psi}^\dag_{n\sigma}$, to rewrite $\hat{H}$ in momentum space as
\begin{eqnarray}
\hat{H}=
\sum_{\mathbf{k}}\Big\{&&\sum_{\sigma}\big[(\alpha_\mathbf{k}+\eta_\sigma h)\hat{\psi}_{\mathbf{k}\sigma }^{\dag}\hat{\psi}_{\mathbf{k}\sigma }+\Delta\hat{\psi}_{\mathbf{k+K}\sigma }^\dag \hat{\psi}_{\mathbf{k}\sigma}\big] \notag\\
&&+\beta_\mathbf{k}\hat{\psi}_{\mathbf{k+K}\uparrow }^\dag\hat{\psi}_{\mathbf{k}\downarrow}-\beta^*_\mathbf{k}\hat{\psi}_{\mathbf{k+K}\downarrow }^\dag\hat{\psi}_{\mathbf{k}\uparrow}\Big\},
\label{eq:Hamk}
\end{eqnarray}
{where $\alpha _{\bf k}=2( \cos k_{x}+\cos k_{y})$ is the identical dispersion relation for both spins hopping in a square lattice, and $\beta_{\mathbf{k}}=2t^{\prime }(i\sin k_{x}+\sin k_{y})$ is a lattice version of spin-orbit coupling \cite{BernevigZhang2006} that couples opposite spins at $\mathbf{k}$ and $\mathbf{k+K}$; $\mathbf{K}=(\pi,\pi)$ is the result of  checkerboard patterns of the staggered potential and the spin-orbit coupling that double the primitive cell of the square lattice (Fig. \ref{fig1}). The details of derivation can be referred to in Appendix \ref{AA}.}

Under the basis $\{{\mathbf{k} \uparrow},{\mathbf{k+K}\uparrow},{\mathbf{k} \downarrow},{\mathbf{k+K}\downarrow}\}$ in Eq. \eqref{eq:Hamk},  the Hamiltonian matrix reads
\begin{equation}
H_{\mathbf{k}}=\left(
\begin{array}{cccc}
\alpha _{\mathbf{k}}+h & \Delta & 0 & -\beta_{\mathbf{k}}\\
\Delta & -\alpha _{\mathbf{k}}+h & \beta_{\mathbf{k}} & 0 \\
0 & \beta_{\mathbf{k}}^{\ast } & \alpha _{\mathbf{k}}-h & \Delta \\
-\beta_{\mathbf{k}}^{\ast } & 0 & \Delta & -\alpha _{\mathbf{k}}-h
\end{array}\right),
\label{eq:h_matrix}
\end{equation}
and the four dressed bands can be obtained and understood in the following three steps:

(1) Spin-degeneracy lifting.
The band degeneracy $\alpha_\mathbf{k}$ for both spins is lifted by the complex Zeeman field $h$ to $\alpha_\mathbf{k}\pm h$ in the complex energy plane, forming the diagonal entries of $H_{\mathbf{k}}$.

(2) Band shifting \& repulsion.
The intra-spin coupling from the staggered potential $\Delta$ shifts parts of the energy band of each spin by $\mathbf{K}$ from the 1st Brillouin zone (BZ) of the square lattice to a smaller one of the checkerboard lattice (see the third panel of Fig. \ref{fig1}) and generally opens complex energy gaps. Thus, four energy bands (called uncoupled bands henceforth) $\{\epsilon_\mathbf{k}^\pm, -\epsilon_\mathbf{k}^\pm\}$ with $\epsilon_\mathbf{k}^\pm=\pm\sqrt{\alpha_\mathbf{k}^2+\Delta^2}+ h$ are formed by diagonalizing both diagonal $2\times 2$ blocks of $H_{\mathbf{k}}$ in the absence of spin-orbit coupling $\beta_{\mathbf{k}}$.

(3) Spin-orbit coupling.
When the spin-orbit coupling $\beta_\mathbf{k}$ is involved, the above diagonalization process block diagonalizes $H_{\mathbf{k}}$ to (see Appendix \ref{AA} for details)
\begin{equation}
H^{b}_{\mathbf{k}}
=\left(
\begin{array}{cccc}
\epsilon^+_{\mathbf{k}} & -\beta_{\mathbf{k}} & 0 & 0 \\
-\beta_{\mathbf{k}}^{\ast } & -\epsilon^+_{\mathbf{k}}  & 0 & 0 \\
0 & 0 & \epsilon^-_{\mathbf{k}} & \beta_{\mathbf{k}} \\
0 & 0 & \beta_{\mathbf{k}}^{\ast } & -\epsilon^-_{\mathbf{k}}
\end{array}\right)
\equiv\left(
\begin{array}{cc}
H^{(1)}_{\mathbf{k}} & 0 \\
0 & H^{(2)}_{\mathbf{k}} \\
\end{array}\right),
\label{eq:block_matrix}
\end{equation}
where each block has the form of a two-band Chern insulator \cite{HaldaneHaldane1988,BernevigZhang2006}, but different from the Hermitian case, the two uncoupled bands, say $\pm\epsilon_\mathbf{k}^-$ of $H^{(2)}_\mathbf{k}$, are generally complex.
Finally, four dressed bands, $\pm d_{\mathbf{k}}^{\pm}\equiv\pm\sqrt{(\epsilon_\mathbf{k}^\pm)^2+|\beta_\mathbf{k}|^2}$ (where two ``$\pm$" symbols are uncorrelated), can be obtained by further diagonalizing each block. 

For Hermitian cases (i.e., $h$ and $\Delta$ are both real numbers), if a spin-orbit coupling with the form of $\beta_{\mathbf{k}}$ is involved, to realize a topologically nontrivial Chern insulator only requires a band inversion of the two uncoupled bands in BZ \cite{BernevigZhang2006}, which requires the model's parameters satisfying $0<h^2-\Delta^2<16$ to reach a topological phase in this model. 
Note that the sign inverse of $h$ only flips spins and thus the sign of Chern number, and that the sign of $\Delta$ doesn't affect the Chern number because it only couples spins of the same species. Thus, in the following, we only focus on the cases of $(h_{r},\Delta_{r})\ge 0$ with subscripts ``$r(i)$" henceforth standing for the real (imaginary) part of corresponding quantities.

\begin{figure}
  \includegraphics[width=1\linewidth]{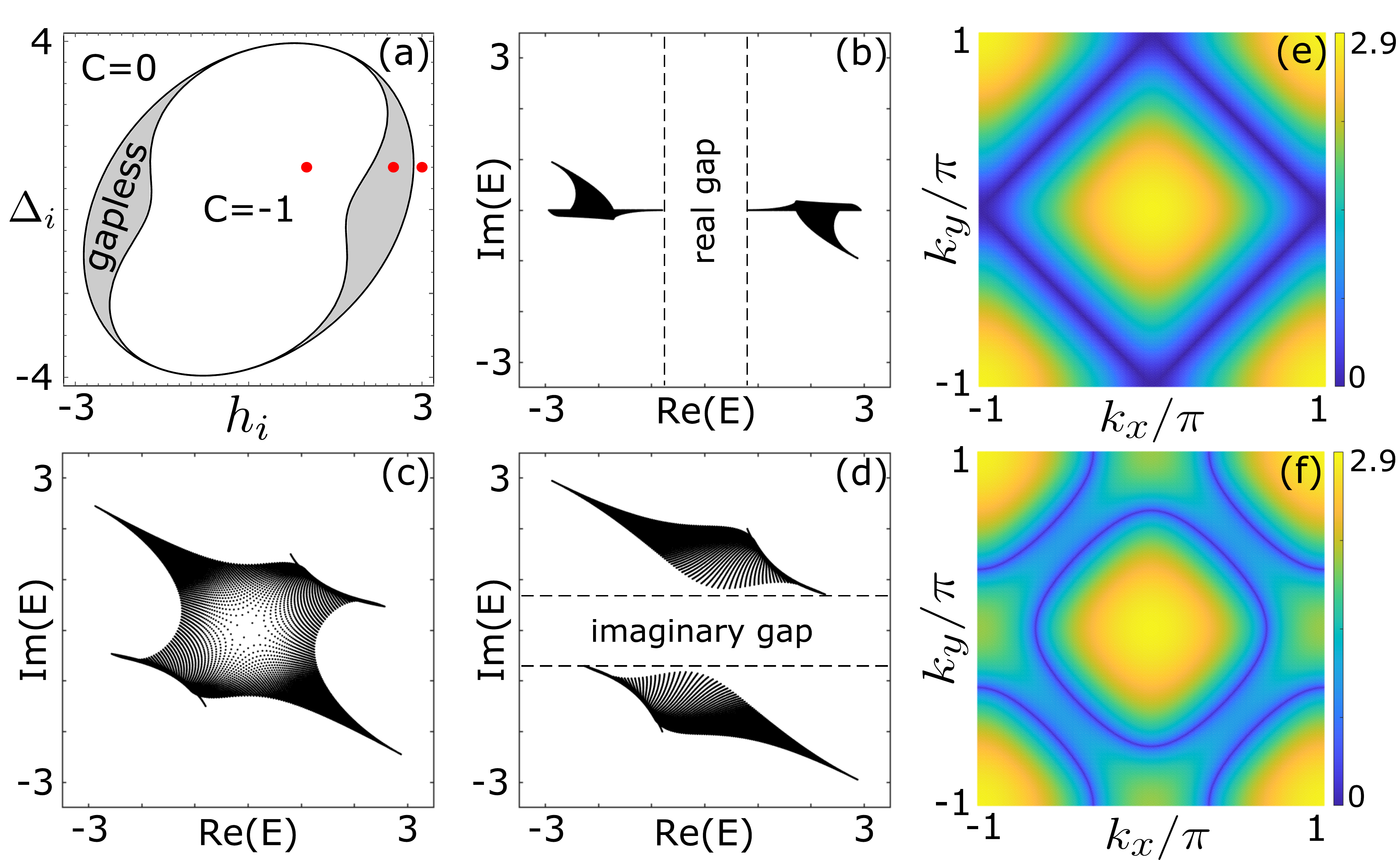}
  \caption{(a) A typical phase diagram calculated by Eq. \eqref{eq:phase_boundary} for $(t',h_r,\Delta_r)=(1,1,0.2)$ with respect to $(h_i,\Delta_i)$, involving three phases: topological ($C=-1$), trivial ($C=0$), and gapless.
  (b-d) Typical energies of $H^{(2)}_\mathbf{k}$ in complex planes for the three phases (denoted by red dots) in (a) with respectively {$(h_i,\Delta_i)=(1,1), (2.5,1)$}, and $(3,1)$. Real and imaginary gaps are shown for the two gapped phases.
  (e,f) Nodal and exceptional loops {(depicted by the absolute value of energy difference)} in lower two dressed bands (sorted by real parts) of $H_\mathbf{k}$ with $\Delta=0$ and $i$, respectively; other parameters are the same as (b).}
\label{fig2}
\end{figure}

We take block $H^{(2)}_\mathbf{k}$ as an example and assume a full complex gap in its energy spectrum. The Chern number for a complex band can be defined as \cite{ShenFu2018}
\begin{eqnarray}
C&\equiv&\frac{1}{2\pi}\int_\text{BZ}F_{xy} \text{d}k_x \text{d}k_y,
\label{eq:Chern_number}
\end{eqnarray}
where $F_{xy}= i\langle  \partial_{k_x} u^{(l)}|\partial_{k_y} u^{(r)}\rangle -(x\leftrightarrow y)$ is the Berry curvature defined with the biorthonormal left/right eigenvectors $|u^{(l,r)}\rangle$ of one complex band of $H^{(2)}_\mathbf{k}$.

To ensure a full complex gap of $H^{(2)}_\mathbf{k}$, we should examine the gap-closing condition,  $d_\mathbf{k}^-=\sqrt{(\epsilon_\mathbf{k}^-)^2+|\beta_\mathbf{k}|^2}=0$, of the two dressed bands $\pm d_\mathbf{k}^-$, yielding (the detailed derivation can be referred to in Appendix \ref{AA})
\begin{eqnarray}
\Big\{
\begin{array}{ccl}
\epsilon_\mathbf{k}^-&=&\pm i|\beta_\mathbf{k}| \\
|\beta_\mathbf{k}|&=&|\Delta_r\Delta_i/h_r-h_i|
\end{array}.
\label{eq:phase_boundary}
\end{eqnarray}
For Hermitian cases (i.e., $h_i=\Delta_i=0$), these equations are reduced to  $\epsilon_\mathbf{k}^-=0$ and	$\beta_\mathbf{k}=0$; the former stands for the crossing of two uncoupled bands $\pm \epsilon_\mathbf{k}^-$, while the latter stands for the possible $\mathbf{k}$'s in BZ that vanish the spin-orbit coupling, i.e., $\Gamma$ point (defined in the third panel of Fig. \ref{fig1}) in this model, which happens to be the position of minimum $\epsilon_{\mathbf{k}}^{-}$ (i.e., the onset of band inversion of two uncoupled bands $\pm \epsilon_{\mathbf{k}}^{-}$). 
Therefore, the judgment of band inversion of two uncoupled bands in spin-orbit coupled Hermitian systems can directly determine the gap-closing condition of the two dressed bands and thus indicate the topological phase transition.
However, for non-Hermitian cases, due to the complexity of the two uncoupled bands $\pm\epsilon_\mathbf{k}^-$, the existence of imaginary part of $\epsilon_\mathbf{k}^{-}$ in $d_\mathbf{k}^{-}$ makes that the gap closing cannot be accomplished by vanishing the spin-orbit coupling $\beta_{\mathbf{k}}$. Thus, the condition becomes Eq. \eqref{eq:phase_boundary}, where the former equation means that it is still {\it necessary} for the crossing in the real part of the two uncoupled bands $\pm\epsilon_\mathbf{k}^{-}$, but the imaginary part should be separated by the spin-orbit coupling $\beta_{\mathbf{k}}$ that is not zero anymore. 
In other words, the band inversion (in the real part of two uncoupled bands) for non-Hermitian cases is a {\it necessary but not sufficient} condition for the topological phase, and one cannot only use the inversion condition for uncoupled complex bands to come into a topological phase transition.

To further anatomize Eq. \eqref{eq:phase_boundary} for non-Hermitian cases, one can find that the second equation determines loops of $\mathbf{k}$'s with centers being located at $\Gamma,M$, or $X$ points in BZ, instead of discrete points (say $\Gamma$ point) for Hermitian cases. Therefore, together with the first equation, {the gap-closing condition becomes more tolerant of the parameter change, that is, the gap-closing point of the Hermitian counterpart expands to an interval, generating gapless phases.}
Figure \ref{fig2}(a) shows a typical phase diagram with respect to the imaginary parts $(h_i,\Delta_i)$, which includes three phases: topological, trivial, and gapless. Chern numbers of fully gapped phases [e.g., Figs. \ref{fig2}(b) and \ref{fig2}(d)] can be calculated according to Eq. \eqref{eq:Chern_number}.
Likewise, Block $H^{(1)}_\mathbf{k}$ can be analyzed in the same way, and of course, the above conclusion is also valid for any two-level spin-orbit coupled non-Hermitian systems.

Back to the original four-band model $H_\mathbf{k}$, the Chern number of multiple dressed bands is just the summation of each one calculated by the corresponding block.
From steps (1) and (2), it is obvious that only one of the blocks can support the band inversion and thus the possible, topologically nontrivial bands. Therefore, the phase diagram derived from Eq. \eqref{eq:phase_boundary} is the same as that of $H_\mathbf{k}$ if considering the Chern number of two dressed bands that belong to different blocks.

\section{Gauge-independent Wilson-loop method for a multiband Chern number}
It has been shown \cite{LangZhou2017} that for Hermitian cases, without the staggered potential $\Delta$, Hamiltonian \eqref{eq:Ham} support nodal loops between lower or upper two dressed bands in BZ, which is protected by the underlying nonsymmorphic symmetries; a finite $\Delta$ breaks the symmetries and thus the nodal loops. 
The proof conducted in the Hermitian context, however, is invalid for non-Hermitian cases due to the complexity of energy bands. 

Alternatively, it can be understood from step (2) that whether the nodal loops from the band shifting break or not is determined by the intra-spin coupling $\Delta$, and the identity of spin-orbit coupling strength $|\beta_\mathbf{k}|$ for both blocks in Eq. \eqref{eq:block_matrix} just {\it preserves} nodal points or gaps. 
Therefore, to generally have nodal loops between ``lower/upper" two dressed bands (sorted by real or imaginary parts), one just needs $\epsilon_{\mathbf{k}}^+=\epsilon_{\mathbf{k}}^-$, i.e., $\alpha_\mathbf{k}^2=-\Delta^2$, which requires that $\Delta$ must be purely imaginary as an extension to non-Hermitian cases.
As a result, a nodal BZ boundary at $\Delta=0$ [Fig. \ref{fig2}(e)] is split into two loops for a purely imaginary $\Delta$ [Fig. \ref{fig2}(f)] according to $\alpha_{\mathbf{k}_l}=\pm \Delta_i$, where $\mathbf{k}_l$ is the loop position in BZ, and only one in the 1st BZ if we note $\alpha_{\mathbf{k}_l+\mathbf{K}}=-\alpha_{\mathbf{k}_l}$. 
These split nodal loops between ``lower/upper" dressed bands respectively with energies $\pm \sqrt{h^2+|\beta_{\mathbf{k}_l}|^2}$ are just \textit{exceptional} loops because of the defectiveness of $H_{\mathbf{k}_l}$ (see the proof in Appendix \ref{AA}).
From $\Delta_i=-4$ to $4$, the exceptional loop in the 1st BZ emerges from $\Gamma$ point, expands to the nodal BZ boundary at $\Delta_i=0$, then bounces back and shrinks, and finally vanishes at $\Gamma$ point again. 
The topology of Weyl exceptional rings in three-dimensional dissipative cold atomic gases has already been theoretically studied \cite{XuDuan2017}.

If a subspace $\mathcal{S}$ of $H_\mathbf{k}$ consists of multiple degenerate complex bands, its Chern number should be calculated via
\begin{eqnarray}
C&\equiv&\frac{1}{2\pi}\int_\text{BZ}\text{tr} ({F}_{xy}) \text{d}k_x \text{d}k_y,
\label{eq:nonabelian}
\end{eqnarray}
where ${F}_{xy}=(\partial_{x} A_y-\partial_{y}A_x) -i[A_x,A_y]$ and $A_{\mu=x,y}$ are respectively non-Hermitian generalizations of the non-Abelian Berry curvature and Berry connection with the component $A^{nm}_{\mu}=i\langle  u_n^{(l)}|\partial_{\mu} u_m^{(r)}\rangle~(\partial_{\mu}\equiv\partial/\partial{k_\mu}$) defined by the biorthonormal {\it dual} left/right eigenvectors $|u^{(l,r)}_{n,m}\rangle$ for bands $n$ and $m$ in the subspace $\mathcal{S}$. 
{We demonstrate in Appendix \ref{AB} that the generalization with \textit{single} left/right eigenvectors is not a proper definition because of the non-covariance of non-Abelian Berry curvature to a unitary transformation; note that the covariance is not required in the Abelian case that is proved to give identical Chern numbers defined with either dual or single left/right eigenvectors \cite{ShenFu2018}.}

To numerically calculate Eq. \eqref{eq:nonabelian}, we develop a gauge-independent method based on a Wilson loop, which is defined for a path from $\mathbf{k}_n$ to $\mathbf{k}_m$ in BZ as follows,
\begin{equation}
 W_{mn}\equiv \mathcal{P}\exp \left(i\int_{\mathbf{k}_n}^{\mathbf{k}_m}\sum_{\mu=x,y}A_\mu dk_\mu\right),
 \label{wilson_line}
\end{equation}
where $\mathcal{P}$ is a path-ordering operator.
Then, we have divided the BZ into many infinitesimal plaquettes $s_j$, and the Chern number $C$ of the subspace $\mathcal{S}$ is just the summation of Chern densities $c_j$ of all plaquettes over the first BZ, i.e., 
\begin{eqnarray}
C=\sum_{s_j\in \text{BZ}}c_j, 
\end{eqnarray}
where
\begin{eqnarray}
c_j = \frac{1}{2\pi i}\ln\big(\det[W^{(j)}_{14}]\det[W^{(j)}_{43}]\det[W^{(j)}_{32}]\det[W^{(j)}_{21}]\big).\notag\\
\label{eq: nAChern}
\end{eqnarray}
The subscripts $\{1,2,3,4\}$ counterclockwise label the four vertices of the plaquette $s_i$ in BZ; $W^{(j)}_{nm}$ is the Wilson line along the plaquette edge  from vertices $m$ to $n$.
{To eliminate numerical errors, we calculate the Chern density using the following formula:
	\begin{eqnarray}
		c_{j}&=&[c^{(cc)}_j-c^{(cl)}_j]/2,
	\end{eqnarray}
	where $c^{(cc)}_j$ and $c^{(cl)}_j$ are the Chern densities calculated respectively using counterclockwise and clockwise Wilson loops.
	The reason why it can eliminate numerical errors is that the Chern densities will be sign-inverted by inverting the Wilson loops, but the errors are accumulated in the same way.}
The detailed derivation can be referred to in Appendix \ref{AB}.

This gauge-independent Wilson-loop method can be regarded as an extension of the Hermitian one \cite{FukuiSuzuki2005}.
Different from Hermitian cases, we should first sort the bands in a proper way (typically by real or imaginary parts) according to the gap types (real or imaginary gaps) for the calculation.
Figure \ref{fig3}(b) shows the consistency of this method with the results of the previous block-diagonalization method. {This method can be used to calculate the Chern number for any number of degenerate complex bands, where the single-band method in Ref. \cite{ShenFu2018} cannot be used anymore because each band cannot be well separated for calculation.}

\section{Boundary-dependent chiral edge states}

\begin{figure}[tb]
  \includegraphics[width=0.9\linewidth]{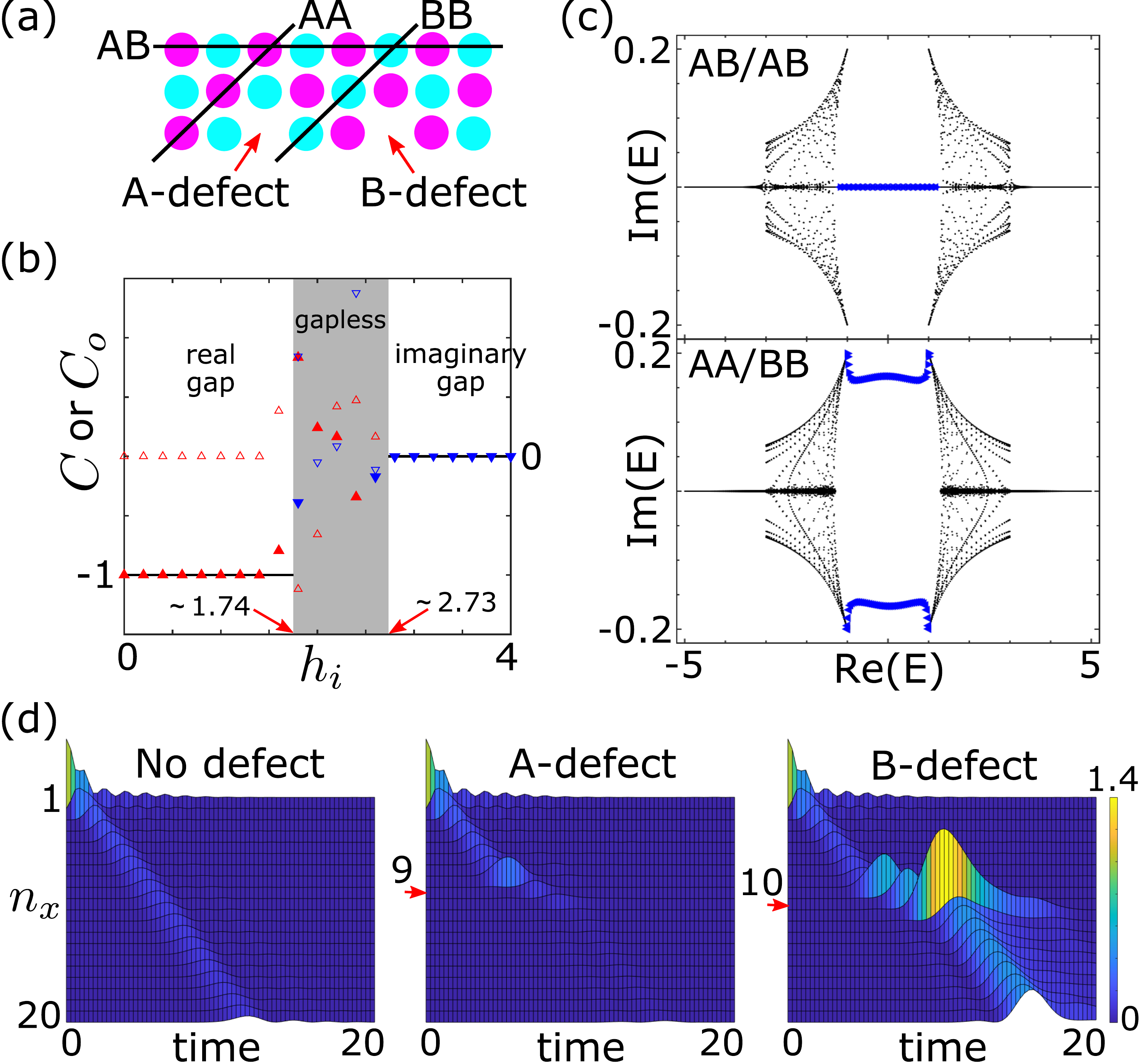}
  \caption{(a) Schematic of different types of open boundaries and defects. 
  (b) Chern number $C$ (solid lines) of Eq. \eqref{eq:nonabelian} under PBCs for the lower two dressed bands by the numerical Wilson-loop method, and open-bulk Chern number $C_o$ (triangles) of Eq. \eqref{eq:open_bulk_Chern} for the subspace $\mathcal{S}$ containing the lowest half number of all states for $N_x=N_y=60$ and $N'_x=N'_y=4$ with four AB boundaries. The shaded ``gapless" region is analytically determined by Eq. \eqref{eq:phase_boundary}. Solid (hollow) triangles represent the real (imaginary) parts of $C_o$; upward(downward)-pointing triangles represent the sorting by real (imaginary) parts of eigenenergies.
  (c) Complex eigenenergies $E$ of a cylinder-shaped lattice with AB/AB boundaries (upper panel) and AA/BB boundaries (lower panel) at the two ends. Triangles represent the edge states in real gaps; left(right)-pointing triangles in the lower panel represent the edge states localized at the AA(BB) boundary with positive(negative) imaginary parts of energies. 
  {(d) Dynamics of the total probability of spin-up and -down states along the bottom AB boundary with no defect (left panel), an A-defect (middle panel), or a B-defect (right panel). A spin-up state (normalized to 1) is initially injected at the bottom-left corner (A site) of a $30\times30$ square lattice with four AB boundaries.}
  All above calculations are done with parameters $(h,\Delta)=(1,0.2i)$ except in (b) where $h_i$ is set as the horizontal axis.}
\label{fig3}
\end{figure}

We also find the reservation of conventional bulk-boundary correspondence by exploiting the open-bulk Chern number under OBCs \cite{210310-1},
\begin{equation}
C_o=\frac{2\pi i}{N'_xN'_y}\text{tr}\big(\hat{P}\big[[\hat{X},\hat{P}],[\hat{Y},\hat{P}]\big]\big),
\label{eq:open_bulk_Chern}
\end{equation} 
where $\hat{X}$($\hat{Y}$) is the coordinate operator for $x(y)$-direction and $\hat{P}=\sum_{s\in \mathcal{S}}|u_s^{(r)}\rangle\langle u_s^{(l)}|$ is the projection operator of subspace $\mathcal{S}$ for the Chern number. The trace is done over a central rectangle $N'_x\times N'_y$ region out of an $N_x \times N_y$ rectangle-shaped lattice. 

Figure \ref{fig3}(b) shows a good match of the open-bulk Chern number $C_o$ under OBCs to the Chern number $C$ under PBCs at gapped regimes if the sorting of eigenenergies is consistent with the bulk-gap type; only ill-defined gapless regime messes up the numerical values.
The equality of the two types of Chern numbers means no non-Hermitian skin effect in this system because the non-Hermiticity only comes from the on-site gain/loss, not from the nonreciprocal hopping \cite{YaoWang2018}.
{Note that although $\Delta$ is set purely imaginary (that guarantees the existence of exceptional loops under PBCs as mentioned above) in Fig. \ref{fig3}, the bulk-boundary correspondence is generally preserved, i.e., $C_o=C$, for all parameter regimes.}

Moreover, different open boundaries give birth to chiral edge states that have different dissipation properties, as shown in Fig. \ref{fig3}(c): The energies are purely real at AB/AB boundaries but complex at AA/BB boundaries {because the gain and loss along the AB boundary are balanced while only a single type of gain or loss exist along AA or BB boundaries;} as a result, the boundary can influence the dynamics of chiral edge states, as shown in Fig. \ref{fig3}(d), where the A/B-defect (minimal incorporation with a different boundary) can decrease/increase the amplitude of edge state along with time. This feature was discovered in a $\mathcal{PT}$-symmetric honeycomb lattice \cite{ZhuChen2018}, but the reality of edge spectra in Fig. \ref{fig3}(c) is robust even though our system lacks this symmetry (i.e., $h_i\ne 0$).

\section{Conclusion and discussion}
In conclusion, we discuss the effects of gain/loss on spin-orbit coupled ultracold atoms in 2D optical lattices, and demonstrate the interplay of non-Hermiticity and the spin-orbit coupling. 
We analytically obtain the topological phase diagrams and unveil that the band inversion is just a necessary but not sufficient condition for a topological phase in two-level spin-orbit coupled non-Hermitian systems.
We also develop a gauge-independent Wilson-loop method for numerically calculating the Chern number of multiple degenerate complex bands. 
Moreover, the conventional bulk-boundary correspondence preserves due to the lack of non-Hermitian skin effect, but the dissipation of chiral edge states under OBCs can be controlled by the boundary selection and thus influences the dynamics of edge states.

{Recently, we have noted that the effect of atom loss (non-Hermiticity) on the dispersion relation of one-dimensional spin-orbit-coupled fermions has been experimentally observed \cite{RenJo2021}. The mature method therein of realizing atom loss in cold atoms \cite{LiLuo2019,LappGadway2019,GouYan2020,TakasuTakahashi2020,FerriEsslinger2021,DingZhang2021,RenJo2021} is also applicable to 2D spin-orbit coupled ultracold atomic systems. 
To experimentally realize the model Hamiltonian \eqref{eq:Ham}, we can take the 2D spin-orbit coupled ultracold systems realized in Ref. \cite{WuPan2016} as the basis, and then add the spin- and site-dependent atom losses by the single near-resonant beams coupling corresponding hyperfine levels \cite{RenJo2021}. Although only atom loss is used, we can deduct the effect of overall loss to realize the relative gain and loss. The loss strength can be tuned by the power of the loss beam.
The spectroscopy for directly measuring complex bands with cold atoms in optical lattices is still a challenge, but the dynamics of edge states is straightforward to observe experimentally if the boundary can be engineered properly.}

\begin{acknowledgments}
L.-J.L.~was supported by the National Natural Science
Foundation of China (Grant No.~11904109), the Guangdong Basic and Applied Basic Research Foundation (Grant No.~2019A1515111101), and the Science and Technology Program of Guangzhou (Grant No.~2019050001);
S.-L.Z.~was supported by the Key-Area Research and Development Program of Guangdong Province (Grant No. 2019B030330001) and the National Natural Science Foundation of China (Grants No. 12074180 and No. U1801661).
\end{acknowledgments}

\appendix

\section{Conditions for non-Hermitian topological phase transitions}
\label{AA}

In the main text, using the discrete Fourier transform in a square lattice with $N$ sites,
$\hat{\psi}^\dag_{\mathbf{k}\sigma}=N^{-1/2}\sum_{n}e^{i\mathbf{k}\cdot \mathbf{R}_n}\hat{\psi}^\dag_{n\sigma}$,
Hamiltonian $\hat{H}$ of Eq. \eqref{eq:Ham} under PBCs in real space can be transformed to Eq. \eqref{eq:Hamk} in momentum space. 
The terms of the intra-spin hopping and the Zeeman field can be dealt with straightforwardly; we only give the derivation for the term of the spin-orbit coupling, and the term of the staggered potential can be likewise obtained.

The derivation for the term of the spin-orbit coupling is as follows:
\begin{widetext}
\begin{eqnarray}
\sum_{\sigma}\sum_{\langle nm \rangle}(-1)^{n_x+n_y}t^{\prime }e^{i\eta_\sigma\phi_{m}}\hat{\psi}_{n\sigma }^\dag\hat{\psi}_{m\bar{\sigma} } 
&=&t'\sum_{\sigma}\sum_{\mathbf{kk}'}\hat{\psi}_{\mathbf{k}'\sigma }^{\dag}\hat{\psi}_{\mathbf{k}\bar{\sigma}}
\Big[\frac{1}{N}\sum_{\langle nm \rangle}(-1)^{n_x+n_y}e^{i\eta_\sigma\phi_{m}}e^{-i(\mathbf{k}'\cdot \mathbf{R}_n-\mathbf{k}\cdot \mathbf{R}_m)}\Big] \notag\\
&=&t'\sum_{\sigma}\sum_{\mathbf{kk}'}\hat{\psi}_{\mathbf{k}'\sigma }^{\dag}\hat{\psi}_{\mathbf{k}\bar{\sigma}}
\Big[\frac{1}{N}\sum_{n}(-1)^{n_x+n_y}e^{-i(\mathbf{k}'-\mathbf{k})\cdot \mathbf{R}_n}\Big]
\sum_{l}e^{i\eta_\sigma\phi_{l}}e^{i\mathbf{k}\cdot \mathbf{R}_l} 
\label{eq:relative} \\
&=&\frac{t'}{2}\sum_{\sigma}\sum_{\mathbf{kk}'}\hat{\psi}_{\mathbf{k}'\sigma }^{\dag}\hat{\psi}_{\mathbf{k}\bar{\sigma}}
\Big[\frac{1}{N_\text{cell}}\big(\sum_{n\in \text{A}}-\sum_{n\in \text{B}}\big)e^{-i(\mathbf{k}'-\mathbf{k})\cdot \mathbf{R}_n}\Big]
\sum_{l}e^{i\eta_\sigma\phi_{l}}e^{i\mathbf{k}\cdot \mathbf{R}_l}
\label{eq:bipart} \\
&=&\frac{t'}{2}\sum_{\sigma}\sum_{\mathbf{kk}'}\hat{\psi}_{\mathbf{k}'\sigma }^{\dag}\hat{\psi}_{\mathbf{k}\bar{\sigma}}
\sum_\mathbf{K}\left(1-e^{-iK_xa}\right)\delta_{\mathbf{k'-k,K}}
\sum_{l}e^{i\eta_\sigma\phi_{l}}e^{i\mathbf{k}\cdot \mathbf{R}_l} 
\label{eq:select} \\
&=&t'\sum_{\mathbf{k}\sigma}\hat{\psi}_{\mathbf{k+K}\sigma }^{\dag}\hat{\psi}_{\mathbf{k}\bar{\sigma}}
\sum_{l}e^{i\eta_\sigma\phi_{l}}e^{i\mathbf{k}\cdot \mathbf{R}_l}
\label{eq:sock}
\end{eqnarray}
\end{widetext}
In Line \eqref{eq:relative}, we use the relative position $\mathbf{R}_l=\mathbf{R}_m-\mathbf{R}_n$; in Line \eqref{eq:bipart}, we divide the square lattice into two checkerboard sublattices A and B with $N_{\text{cell}}=N/2$ primitive cells, and the origin is set at an A site and the relative position of a B site to the A site in the same primitive cell is $a\hat{\mathbf{x}}$; in Line \eqref{eq:select}, we use the selection rule
\begin{eqnarray}
\frac{1}{N_\text{cell}}\sum_{n\in \text{A}}e^{i\mathbf{k}\cdot \mathbf{R}_n}
=\sum_\mathbf{K}\delta_{\mathbf{k,K}},
\label{eq:selrul}
\end{eqnarray}
where $\mathbf{K}$ is a reciprocal lattice vector of the checkerboard lattice A,  i.e., $\mathbf{K}\cdot\mathbf{R}_{n\in \text{A}}=2\pi \times$integer; because $\mathbf{k}$ is a vector in the 1st BZ of the square lattice, the selection rule only requires that $\mathbf{K}=(0,0)$ and $(\pi,\pi)/a$, yielding Line \eqref{eq:sock}.
Considering the specific phase values of $\phi_l$ in the main text, we can get the last two terms in Eq. \eqref{eq:Hamk}.

Using the basis $\{{\mathbf{k} \uparrow},{\mathbf{k+K}\uparrow},{\mathbf{k} \downarrow},{\mathbf{k+K}\downarrow}\}$ in Eq. \eqref{eq:Hamk},  the Hamiltonian matrix $H_\mathbf{k}$ of Eq. \eqref{eq:h_matrix} can be block diagonalized as follows:
\begin{widetext}
\begin{eqnarray}
V^{-1}_\mathbf{k} H_\mathbf{k}V_\mathbf{k}&\equiv&
\left(
\begin{array}{cc}
U^{-1} & 0 \\
0 & U^{-1} 
\end{array}\right)
\left(
\begin{array}{cccc}
\alpha _{\mathbf{k}}+h & \Delta & 0 & -\beta_{\mathbf{k}}\\
\Delta & -\alpha _{\mathbf{k}}+h & \beta_{\mathbf{k}} & 0 \\
0 & \beta_{\mathbf{k}}^{\ast } & \alpha _{\mathbf{k}}-h & \Delta \\
-\beta_{\mathbf{k}}^{\ast } & 0 & \Delta & -\alpha _{\mathbf{k}}-h
\end{array}\right)
\left(
\begin{array}{cc}
U & 0 \\
0 & U 
\end{array}\right)
=\left(
\begin{array}{cccc}
\epsilon_{\mathbf{k}}^+ & 0 & 0 & -\beta_\mathbf{k} \\
0 &\epsilon_{\mathbf{k}}^- & \beta_\mathbf{k} & 0 \\
0 & \beta^*_\mathbf{k} &-\epsilon_{\mathbf{k}}^-& 0 \\
-\beta_\mathbf{k}^* & 0 & 0 &-\epsilon_{\mathbf{k}}^+
\end{array}\right),
\end{eqnarray}
\end{widetext}
where $\epsilon_\mathbf{k}^\pm=\pm\sqrt{\alpha_\mathbf{k}^2+\Delta^2}+h$ and $U=\exp(-i\sigma_y\omega/2)$ is a similarity matrix; $\omega$ is a complex angle defined by the parameterization $(\alpha_\mathbf{k},\Delta)=\sqrt{\alpha_\mathbf{k}^2+\Delta^2}(\cos\omega,\sin\omega)$; other quantities are defined the same as in the main text. By rearranging the basis, we get the block-diagonal matrix $H^b_\mathbf{k}$ in Eq. \eqref{eq:block_matrix}. 
Here, it is worth noting that when $\alpha^2_\mathbf{k}+\Delta^2= 0$, i.e., $\alpha_\mathbf{k}=\pm\Delta_i$, the parameterization fails because the two diagonal $2\times 2$ blocks of $H_\mathbf{k}$ become defective (i.e., non-diagonalizable), but we can still use this parameterization infinitely close to this point; at this point, $H_\mathbf{k}$ also becomes defective because it is mathematically similar to a Jordan canonical form, i.e.,
\begin{widetext}
\begin{eqnarray}
\left(
\begin{array}{cccc}
\pm i\Delta+h & \Delta & 0 & -\beta_{\mathbf{k}}\\
\Delta & \mp i\Delta+h & \beta_{\mathbf{k}} & 0 \\
0 & \beta_{\mathbf{k}}^{\ast } & \pm i\Delta-h & \Delta \\
-\beta_{\mathbf{k}}^{\ast } & 0 & \Delta & \mp i\Delta-h
\end{array}\right)
\sim
\left(
\begin{array}{cccc}
\sqrt{h^2+|\beta_\mathbf{k}|^2} & 1 & 0 & 0 \\
0 & \sqrt{h^2+|\beta_\mathbf{k}|^2} & 0 & 0 \\
0 & 0 & -\sqrt{h^2+|\beta_\mathbf{k}|^2} & 1 \\
0 & 0 & 0 & -\sqrt{h^2+|\beta_\mathbf{k}|^2}
\end{array}\right), \notag\\
\end{eqnarray}
\end{widetext}
which means that the nodal loops between ``upper/lower" two dressed bands, whose position $\mathbf{k}_l$ in BZ is determined by $\alpha_{\mathbf{k}_l}=\pm\Delta_i$, are just  {\it exceptional} loops  with energies $\pm \sqrt{h^2+|\beta_{\mathbf{k}_l}|^2}$.

Taking block $H^{(2)}_\mathbf{k}$ in Eq. \eqref{eq:block_matrix} as an example, its two eigenenergies are 
$\pm\sqrt{(\epsilon_\mathbf{k}^-)^2+|\beta_\mathbf{k}|^2}\equiv \pm d$. The complex-gap closing condition requires that $d=0$, that is, 
\begin{eqnarray}
&\epsilon_\mathbf{k}^-=-\sqrt{\alpha_\mathbf{k}^2+\Delta^2}+ h=\pm i|\beta_\mathbf{k}|& \notag\\
\Rightarrow ~~& \alpha_\mathbf{k}^2+\Delta^2=(h\pm i|\beta_\mathbf{k}|)^2,&
\end{eqnarray}
which, considering the real and imaginary parts separately, can be reexpressed by real parameters  as
\begin{eqnarray}
\Big\{
\begin{array}{rcl}
\alpha_\mathbf{k}^2+\Delta_r^2-\Delta_i^2&=&h_r^2-(h_i\pm |\beta_\mathbf{k}|)^2 \\
\Delta_r\Delta_i&=&h_r(h_i\pm |\beta_\mathbf{k}|)
\end{array}.
\end{eqnarray}
Solving these simultaneous equations for $\alpha_\mathbf{k}^2$ and $|\beta_\mathbf{k}|^2$, we have ($t=a=1$)
\begin{eqnarray}
\Big\{
\begin{array}{ccl}
\alpha_\mathbf{k}^2&=&4(\cos k_x+\cos k_y)^2=(h_r^2-\Delta_r^2)(1+\Delta_i^2/h_r^2) \\
|\beta_\mathbf{k}|^2&=&4t'^2(\sin^2 k_x+\sin^2 k_y)=(\Delta_r\Delta_i/h_r-h_i)^2
\end{array}.
\label{seq:condition}
\end{eqnarray}
We can also reexpress them in terms of $\epsilon_\mathbf{k}^-$ and $\beta_\mathbf{k}$ in $H^{(2)}_\mathbf{k}$ as
\begin{eqnarray}
	\Big\{
	\begin{array}{ccl}
		\epsilon_\mathbf{k}^-&=&\pm i|\beta_\mathbf{k}| \\
		|\beta_\mathbf{k}|&=&|\Delta_r\Delta_i/h_r-h_i|
	\end{array},
\end{eqnarray}
which is just Eq. \eqref{eq:phase_boundary} in the main text.

Note that for the simultaneous equations in \eqref{seq:condition}, given all the Hamiltonian parameters, the first equation determines a loop with the center located at $\Gamma$ point (defined in the third panel of Fig. \ref{fig1}) in the 1st BZ, and the second one determines loops with centers located at $\Gamma$ and $M$ points or at $X$ points (also defined in the third panel of Fig. \ref{fig1}) in the 1st BZ. Therefore, the solutions to Eq. \eqref{seq:condition} are just the intersection of two loops from different equations. 
The touch of the two loops along with the change of Hamiltonian parameters means the phase transition between a gapped phase and a gapless phase. 
Because of the $C_4$ symmetry of each equation in Eq. \eqref{seq:condition}, the touch points can only happen along the lines of $k_y=\pm k_x$ or of $k_{x,y}=0$ in the 1st BZ, using which the phase boundaries between gapped and gapless phases can be determined by 
\begin{eqnarray}
&&\Big[\sqrt{(h_r^2-\Delta_r^2)(1+\Delta_i^2/h_r^2)}-2\Big]^2 \notag\\
&&~~~~~~~~~~~~~~~~ +(\Delta_r\Delta_i/h_r-h_i)^2/t'^2=4
\end{eqnarray}
and 
\begin{eqnarray}
(h_r^2-\Delta_r^2)(1+\Delta_i^2/h_r^2)+2(\Delta_r\Delta_i/h_r-h_i)^2/t'^2&=&16. \notag\\
\end{eqnarray}

\section{Properties of the Wilson-loop method}
\label{AB}

For convenience, we first define matrices for sets of $M$ interested right/left eigenvectors, $\{|u^{(r,l)}_s\rangle\}~(s=1,\cdots,M)$, of an ${N\times N}$ Hamiltonian matrix $H(\mathbf{k})$ with a two-dimensional parameter $\mathbf{k}$, 
\begin{equation}
\Theta^{(r,l)\dag}=(\{|u_s^{(r,l)}\rangle \})_{N\times M}.
\end{equation}
The bi-orthonormality of right/left eigenvectors requires
\begin{equation}
 \Theta^{(l)}\Theta^{(r)\dag}=I, 
 \label{eq:biorthonormal}
\end{equation}
where $I$ is an ${M\times M}$ identity matrix, and a projector operator for a subspace $\mathcal{S}$ can be defined as
\begin{equation}
\hat{P}= \Theta^{(r)\dag}\Theta^{(l)}=\sum_{s\in \mathcal{S}}|u_s^{(r)}\rangle\langle u_s^{(l)}|, 
\end{equation}
which is an $N\times N$ matrix.

Using the differential form notation, the 1-form of the non-Hermitian non-Abelian Berry connection is defined as
\begin{equation}
A\equiv \sum_{\mu=x,y}A_\mu dk_\mu, 
\end{equation}
where
$A_\mu\equiv i\Theta^{(l)}\partial_\mu \Theta^{(r)\dag}
=-i(\partial_\mu\Theta^{(l)})\Theta^{(r)\dag}$ 
is an $M\times M$ matrix with $\partial_\mu\equiv\partial/\partial{k_\mu}$.

The 2-form of the non-Hermitian non-Abelian Berry curvature is defined as
\begin{eqnarray}
F&\equiv&dA-iA\wedge A 
=F_{xy} dk_xdk_y,
\end{eqnarray}
where 
\begin{equation}
F_{xy}=i[\partial_{x} \Theta^{(l)}]\hat{O}[\partial_{y} \Theta^{(r)\dag}]-({x}\leftrightarrow{y})
=-F_{yx}
\end{equation} 
is an $M\times M$ matrix, and $\hat{O}\equiv 1-\hat{P}$ is the projection operator (an $N\times N$ matrix) for the subspace complementary to $\mathcal{S}$.
Note that $\mathrm{tr}(F)=\mathrm{tr}(dA)$.

The Chern number for multiple degenerate complex bands can be defined as 
\begin{eqnarray}
C&=&\sum_{s_j\in \text{BZ}}c_j
\equiv\sum_{s_j\in \text{BZ}}\frac{1}{2\pi}\int_{s_j} {\rm tr} (F) \notag\\
&=&\sum_{s_j\in \text{BZ}}\frac{1}{2\pi}\int_{s_j}\mathrm{tr}(dA) 
=\sum_{s_j\in \text{BZ}}\frac{1}{2\pi}\oint_{\partial s_j}\mathrm{tr}(A),
\label{eq: nAChern}
\end{eqnarray}
where we have divided the BZ into many infinitesimal plaquettes $s_j$, and the Chern number $C$ is the summation of Chern densities $c_j$ of all plaquettes over the whole BZ.
In the third identity, $\mathrm{tr}(A\wedge A)=0$ is used, and in the last identity, the use of Stokes' theorem transforms an integral over the plaquette surface $s_j$ to an integral through the closed boundary of $s_j$, denoted by $\partial s_j$.

The Wilson line following a path from $\mathbf{k}_n$ to $\mathbf{k}_m$ in BZ can be defined as follows,
\begin{equation}
 W_{mn}\equiv \mathcal{P}\exp \Big(i\int_{\mathbf{k}_n}^{\mathbf{k}_m}A\Big),
 \label{wilson_line}
\end{equation}
which is also an $M\times M$ matrix, and where $\mathcal{P}$ is a path-ordering operator acting on the matrix.
For an infinitesimal Wilson line $\mathbf{k}_m\rightarrow\mathbf{k}_n$, we have
\begin{eqnarray}
&W_{mn}= e^{iA}\approx I+iA
= \Theta^{(l)}_{m}\Theta^{(r)\dag}_{n}, &\notag\\
&W_{nm}=W_{mn}^{-1}\approx \Theta^{(l)}_{n}\Theta^{(r)\dag}_{m},&
\label{eq:infinitesimal_W}
\end{eqnarray}
and thus, the Chern density $c_j$ can be numerically calculated as follows: 
\begin{eqnarray}
c_j&=& \frac{1}{2\pi i}\ln\big(\det[W^{(j)}_{14}]\det[W^{(j)}_{43}]\det[W^{(j)}_{32}]\det[W^{(j)}_{21}]\big) \notag\\
&=&\frac{1}{2\pi i}\ln \exp\Big[i\oint_{\partial s_j}\mathrm{tr}(A)\Big]
=\frac{1}{2\pi}\oint_{\partial s_j} \mathrm{tr}(A) 
\label{eq:equiv} \\
&\approx&\frac{1}{2\pi i}\ln\big(\det[\Theta^{(l)}_{1}\Theta^{(r)\dag}_{4}]\det[\Theta^{(l)}_{4}\Theta^{(r)\dag}_{3}] \notag\\
&&~~~~~~~~~~~~\times\det[\Theta^{(l)}_{3}\Theta^{(r)\dag}_{2}]\det[\Theta^{(l)}_{2}\Theta^{(r)\dag}_{1}]\big)
\label{eq:numerical_chern}
\end{eqnarray}
where $\{1,2,3,4\}$ counterclockwise label the four vertices of the plaquette $s_i$ in the BZ; $W^{(j)}_{nm}$ is the Wilson line along the plaquette edge  from vertices $m$ to $n$. In principle, the plaquette must be small enough such that each Chern density satisfies $|c_j| \le 1/2$ to ensure that the $\ln$ operation does not miss some part of the value due to the showing up of $c_j$ as a phase modulo $2\pi$. 
Eq. \eqref{eq:equiv} shows the exact equivalence to the analytical result Eq. \eqref{eq: nAChern} when the plaquettes are infinitely small, where we use the relation
\begin{eqnarray}
\det[W_{mn}^{(j)}]&=&\det(e^{iA})=e^{i\mathrm{tr}(A)}.
\end{eqnarray}

To eliminate the numerical errors from the approximation \eqref{eq:numerical_chern}, we can use the trick by simultaneously calculating the Chern density with the Wilson loop {\it clockwise} along each plaquette edge, yielding
\begin{eqnarray}
c_{j}&=&[c^{(cc)}_j-c^{(cl)}_j]/2,
\end{eqnarray}
where $c^{(cc)}$ and $c^{(cl)}$ are Chern densities calculated by counterclockwise and clockwise Wilson-loop schemes, respectively.
The reason why it can eliminate the numerical errors is that the Chern densities will be sign-inverted by inverting the Wilson loops, but the errors are accumulated in the same way.

In the following, we show the dependence of the above quantities on a similarity transformation. 

Consider a similarity transformation for left/right eigenstates as follows:
\begin{equation}
\tilde{\Theta}^{(l)}=R^{-1}\Theta^{(l)},~~~\tilde{\Theta}^{(r)\dag}=\Theta^{(r)\dag}R.
\end{equation}
which conserves the bi-orthonormality, i.e.,
\begin{equation}
\tilde{\Theta}^{(l)}\tilde{\Theta}^{(r)\dag}=\Theta^{(l)}\Theta^{(r)\dag}=I.
\end{equation}
Under this transformation, the Berry connection is not covariant, because
\begin{equation}
\tilde{A}= \sum_{\mu=x,y}\tilde{A}_\mu dk_\mu
=R^{-1}AR+iR^{-1}dR,
\end{equation}
where $\tilde{A}_\mu=i\tilde{\Theta}^{(l)}\partial_\mu \tilde{\Theta}^{(r)\dag}$,
but the Berry curvature is covariant:
\begin{eqnarray}
\tilde{F}&\equiv &d\tilde{A}-i\tilde{A}\wedge \tilde{A} 
=R^{-1}FR,
\end{eqnarray}
and thus, the trace and the determinant are both invariant, i.e., 
\begin{eqnarray}
\mathrm{tr}{(\tilde{F})}=\mathrm{tr}{(F)},~~~\mathrm{det}{(\tilde{F})}=\mathrm{det}{(F)}.
\end{eqnarray}
According to Eq. \eqref{eq:infinitesimal_W}, a Wilson line is covariant to the transformation because
\begin{equation}
 \tilde{W}_{mn}= R_{m}^{-1}WR_{n},
\end{equation}
and thus the Wilson loop (i.e., $m=n$) is gauge-independent.
So, our numerical method to calculate the Chern number \eqref{eq:numerical_chern} based on the Wilson loops is {\it gauge-independent}, which can be regarded as an extension of the Hermitian method in Ref. \cite{FukuiSuzuki2005} to the non-Hermitian regime. 
In Ref. \cite{HouZhang2021}, the authors use another symmetric definition with dual left/right eigenvectors, but it is obvious that the Wilson loop is not gauge-independent because of $R^{-1}\ne R^\dag.$

We should note that this definition of Chern number for multiple degenerate complex bands with respect to dual left/right eigenvectors is in principle only valid for a set of bands {\it without} exceptional points in between, because left eigenvectors cannot be well defined at these points. However, we can avoid selecting them for numerical calculation when they are just several discrete exceptional points. 
The numerical calculation works well, but actually, we haven't proved this, and a further question is, for an unavoidable bunch of exceptional points, e.g., exceptional surface, how to do the calculation, which deserves future studies.

If we use single right/left eigenvectors to define the above quantities, we have found the differences as follows (for brevity we omit the superscript $(r,l)$):

(1) The bi-orthonormal condition in Eq. \eqref{eq:biorthonormal} becomes a normal but nonorthogonal condition:
\begin{equation}
 \Theta\Theta^{\dag}=J, 
 \label{eq:nonorthogonal}
\end{equation}
where $J$ is an ${M\times M}$ matrix with diagonal entries being 1's and non-zero $\mathbf{k}$-dependent off-diagonal entries due to the non-orthogonality for different right eigenvectors.
And thus, $A_\mu\equiv i\Theta\partial_\mu \Theta^{\dag}
=-i(\partial_\mu\Theta)\Theta^{\dag}+i\partial_\mu J $.

(2) The infinitesimal Wilson line cannot be expressed in a similar form as in Eqs. \eqref{eq:infinitesimal_W}, but 
\begin{eqnarray}
&W_{mn}\approx (I-J)+\Theta_{m}\Theta^{\dag}_{n}, &\notag\\
&W_{nm}=W_{mn}^{-1}\approx (I-J)+ \Theta_{n}\Theta^{\dag}_{m},&
\end{eqnarray}
which are more complicated than Eqs. \eqref{eq:infinitesimal_W}.
And thus, the expression in Eq. \eqref{eq:numerical_chern} must be changed accordingly.

(3) To ensure the normal but nonorthogonal condition Eq. \eqref{eq:nonorthogonal}, the similarity transformation $R$ should be changed to a unitary transformation $R^\dag=R^{-1}$. 
However, we can verify that both $A$ and $F$ are not covariant to this transformation, which is why we cannot use the single right/left eigenvectors to define the Chern number of multiple degenerate complex bands.

\bibliography{ref}
\bibliographystyle{apsrev4-1}

\end{document}